\documentclass[twocolumn,english]{revtex4-1}
\usepackage[T1]{fontenc}
\usepackage[latin9]{inputenc}
\usepackage{textcomp}
\usepackage{amsmath}
\usepackage{graphicx}



\usepackage{babel}
\begin{document}

\title{Quantum thermodynamic cycle with quantum phase transition }

\author{Yu-Han Ma,$^{1,2}$ Shan-He Su,$^{1}$ Chang-Pu Sun$^{1,2}$}
\email{cpsun@csrc.ac.cn}

\address{$^{1}$Beijing Computational Science Research Center, Beijing 100193,
China~\\
 $^{2}$Graduate School of Chinese Academy of Engineering Physics,
Beijing 100084, China}
\begin{abstract}
With the Lipkin-Meshkov-Glick (LMG) model as an illustration, we construct
a thermodynamic cycle composed of two isothermal processes and two
isomagnetic field processes and study the thermodynamic performance
of this cycle accompanied by the quantum phase transition (QPT). We
find that for a finite particle system working below the critical
temperature, the efficiency of the cycle is capable of approaching
the Carnot limit when the external magnetic field $\lambda_{1}$ corresponding
to one of the isomagnetic processes reaches the crosspoint of the
ground states' energy level, which can become critical point of the
QPT in large $N$ limit. Our analysis proves that the system's energy
level crossings at low temperature limits can lead to significant
efficiency improvement of the quantum heat engine. In the case of
the thermodynamics limit $\left(N\text{\textrightarrow}\text{\ensuremath{\infty}}\right)$,
analytical partition function is obtained to study the efficiency
of the cycle at high and low temperature limits. At low temperatures,
when the magnetic fields of the isothermal processes are located on
both sides of the critical point of the QPT, the cycle obtains the
maximum efficiency and the Carnot efficiency can be achieved. This
observation demonstrate that the QPT of the LMG model below critical
temperature is beneficial to the thermodynamic cycle's operation.
\end{abstract}
\maketitle

\section{Introduction}

A heat engine is a machine that allows the working substance to draw
heat from a heat bath through a thermodynamic cycle and to use part
of the extracted energy to do work. Carnot's theorem states that the
efficiency of a reversible heat engine operating between two heat
baths has a maximum efficiency, which is known as the Carnot efficiency
and serves as an upper bound on the efficiency of any irreversible
heat engines running between the same heat baths. In recent years,
people have paid attention to the circumstances that the working substances
have quantum behaviors. The heat engines, which adopt the quantum
system as the working substance, are called the quantum heat engine
\cite{key-13,key-14,key-3,key-15,key-16}. One particular example
in studying the quantum heat engine is to get a link between the definitions
of heat and work and the microscopic states of the working substances
\cite{key-12,key-1,key-2}. These allow us to calculate the efficiency
of a thermodynamic cycle from a more fundamental perspective. Different
working substances such as harmonic oscillators \cite{key-h1,key-h2,key-h3},
spin systems \cite{key-s1,key-s2}, and multi-level atoms \cite{key-1,key-t1,key-t2}
have been studied to construct quantum heat engines. The basic motivation
for designing quantum heat engines is to use the quantum properties
of working substances or heat baths to improve the engines' efficiencies
\cite{key-4,key-5}.

If the working substances consist of quantum particles with interaction,
the quantum phase transition (QPT) may occur below the critical temperature
due to the continuous tuning of a specific external parameter. The
ground state of the system will vary enormously due to the QPT \cite{key-9,key-19}.
An interesting question arises here is that whether or not the QPT
of the working substance is capable of improving the efficiency of
the quantum heat engine? This question inspires us to construct a
quantum heat engine by taking advantage of the QPT effect in the Lipkin-Meshkov-Glick
(LMG) model \cite{key-20,key-21,key-22,key-18}. The LMG has been
studied in different experimental systems, such as platforms with
Bose-Einstein condensates \cite{key-a,key-b,key-c,key-e} and with
trapped ions \cite{key-f,key-g}. We will prove that when the QPT
of the LMG model takes place below the critical temperature, the efficiency
of the quantum heat engine is improved and will achieve the Carnot
efficiency.

This paper is organized as follows: In Sec. \ref{sec:Lipkin-Meshkov-Glick-Model},
we diagonalize the Hamiltonian of the LMG model and analyze its energy
spectra with level crossing. In Sec. \ref{sec:Entropy-of-the}, the
entropy of the LMG model in the thermal equilibrium state is calculated.
The thermodynamic cycle constructed by the LMG model and the efficiency
expression of this cycle are given in Sec. \ref{sec:Quantum-heat-engine}.
In Sec. \ref{sec:Maximum-efficiency-of}, we carefully analyze the
efficiency of the heat engine with finite number of particles, and
give the efficiency of the heat engine at the crosspoints of the ground
states' energy level. In Sec. \ref{sec:Efficiency-of-the}, we calculate
the analytical expression of the efficiency of the cycle in different
cases when the particle number is taking the thermodynamic limit.
Conclusions are given in Sec. \ref{sec:conclusion}

\section{Lipkin-Meshkov-Glick Model\label{sec:Lipkin-Meshkov-Glick-Model}}

\subsection{Diagonalization of the Hamiltonian}

We suppose that the working substance of the quantum heat engine is
a spin system with interaction, which can be described by the Lipkin-Meshkov-Glick
(LMG) model \cite{key-20,key-21,key-22}. Let $J_{\alpha}=\frac{1}{2}\sum_{i=1}^{N}\sigma_{\alpha}^{i}$
be the total spin of the system, where $\sigma_{i}^{\alpha}\left(\alpha\in\left\{ x,y,z\right\} \right)$
are the Pauli operators of the $i$th spin. The Hamiltonian of the
LMG model reads

\begin{equation}
H=\varepsilon J_{z}+V\left(J_{x}^{2}-J_{y}^{2}\right)+W\left(J^{2}-J_{z}^{2}\right),
\end{equation}
where $J^{2}=\sum_{\alpha=x,y,z}J_{\alpha}^{2}$, $V\textrm{ and }W$
are the parameters describing the interaction strengths between spins,
and $\varepsilon$ is the intensity of the external magnetic field.
Now, by defining $\lambda=-\varepsilon/2$, $\lambda_{0}/N=-\left(W+V\right)/2$,
and setting $\gamma=\left(W-V\right)/(W+V)=1$ \cite{key-22,key-18,key-redifination},
we can simplify the Hamiltonian as

\begin{equation}
H=-\frac{2\lambda_{0}}{N}\left(J^{2}-J_{z}^{2}\right)-2\lambda J_{z}.
\end{equation}
The ground state of $H$ lies in the subspace spanned by $\left\{ \left|N/2,M\right\rangle ,M\in\left[-N/2,N/2\right]\right\} $.
$M$ is restricted to an integer for even $N$. For odd $N$, $M$
is half-integer between $-N/2$ and $N/2$. The state $\left|N/2,M\right\rangle $
satisfies

\begin{equation}
J_{z}\left|\frac{N}{2},M\right\rangle =M\left|\frac{N}{2},M\right\rangle ,
\end{equation}
and

\begin{equation}
J^{2}\left|\frac{N}{2},M\right\rangle =\frac{N}{2}\left(\frac{N}{2}+1\right)\left|\frac{N}{2},M\right\rangle .
\end{equation}
By taking $\lambda_{0}=1$, the eigenenergy corresponding to $\left|N/2,M\right\rangle $
is explicitly obtained as

\begin{equation}
E\left(M\right)=\frac{2}{N}\left(M-\frac{N\lambda}{2}\right)^{2}-\frac{N}{2}\left(1+\lambda^{2}\right)-1\text{.}\label{eq:eigenenergy}
\end{equation}
Obviously, the minimum value of $E\left(M\right)$ is at $M=N\lambda/2$,
so that the ground state of the LMG model is $\lambda$ dependent,
i.e.,

\begin{equation}
\left|G\right\rangle =\begin{cases}
\left|\frac{N}{2},\frac{N}{2}\right\rangle  & \left(\lambda>1\right)\\
\left|\frac{N}{2},I\left(\lambda\right)\right\rangle  & \left(0\text{<}\lambda<1\right)
\end{cases}.\label{eq:ground state}
\end{equation}
 $I\left(\lambda\right)$ is the integer or half-integer nearest to
$N\lambda/2$. Equation (\ref{eq:ground state}) indicates that the
quantum phase transition (QPT) arises at the critical point $\lambda_{c}=1$
\cite{key-18}.

\subsection{Energy level crossing of the ground states}

It follows from Eq. (\ref{eq:ground state}) that the ground state
of the LMG model is $\left|N/2,I\left(\lambda\right)\right\rangle $
for $0<\lambda<1$. If $N\lambda/2=k+\frac{1}{2}$ is a half-integer
for $k=0,1,2,\cdots$. Equation (\ref{eq:eigenenergy}) indicates
that the states $\left|N/2,M=k\right\rangle $ and $\left|N/2,M=k+1\right\rangle $
have the same eigenenergy and both of them are the ground states.
Therefore, when $\lambda\left(k\right)=\left(2k+1\right)/N$, the
ground state is degenerate. Because the energy level crossing of the
ground states only appears when $\lambda<1$, the integer $k\leq\left(N-2\right)/2.$
For even $N$, the number of the crosspoints equals $N/2$. The last
crosspoint of the ground state is at $\lambda\left(k=N/2-1\right)=1-1/N$.
In the thermodynamic limit ($N$$\rightarrow\infty$),

\begin{equation}
\underset{N\rightarrow\infty}{lim}\lambda\left(k=N/2-1\right)=1=\lambda_{c},
\end{equation}
which means that the last crosspoint of the energy levels of the ground
states is exactly the critical point for QPT.

\section{Entropy of the working substance\label{sec:Entropy-of-the}}

To calculate the heat exchange in the isothermal processes of a thermodynamic
cycle, we need to discuss the entropy of the working substance. Considering
the LMG model in thermal contact with a heat bath with temperature
$T_{0}$, the density matrix operator in the canonical ensemble can
be written as

\begin{equation}
\rho=\frac{1}{Z}\sum_{M=-N/2}^{N/2}e^{-\beta_{0}E\left(M\right)}\left|\frac{N}{2},M\right\rangle \left\langle \frac{N}{2},M\right|.\label{eq:desity}
\end{equation}
Here, $Z=\sum_{M}\exp\left[-\beta_{0}E\left(M\right)\right]$ is the
partition function and $\beta_{0}=1/T_{0}$ is the inverse temperature,
as we take the Boltzmann constant $k_{B}=1$. By using Eq. (\ref{eq:desity}),
the entropy of the working substance is obtained as

\begin{equation}
S=-Tr\left(\rho\ln\rho\right)=-\sum_{M=-N/2}^{N/2}\frac{e^{-\beta_{0}E\left(M\right)}}{Z}\ln\frac{e^{-\beta_{0}E\left(M\right)}}{Z}.\label{eq:entropy}
\end{equation}
 According to Eq. (\ref{eq:eigenenergy}), when $\lambda\left(k\right)=(2k+1)/N$,
the states $\left|N/2,M=k\right\rangle $ and $\left|N/2,M=k+1\right\rangle $
have the same energy and are degenerate. This implies that the entropy
of the system at $T=0$ has a sudden abrupt rise at the energy level
crossing.

\section{Quantum heat engine based on the LMG model \label{sec:Quantum-heat-engine}}
\begin{center}
\begin{figure}
\includegraphics[scale=0.25]{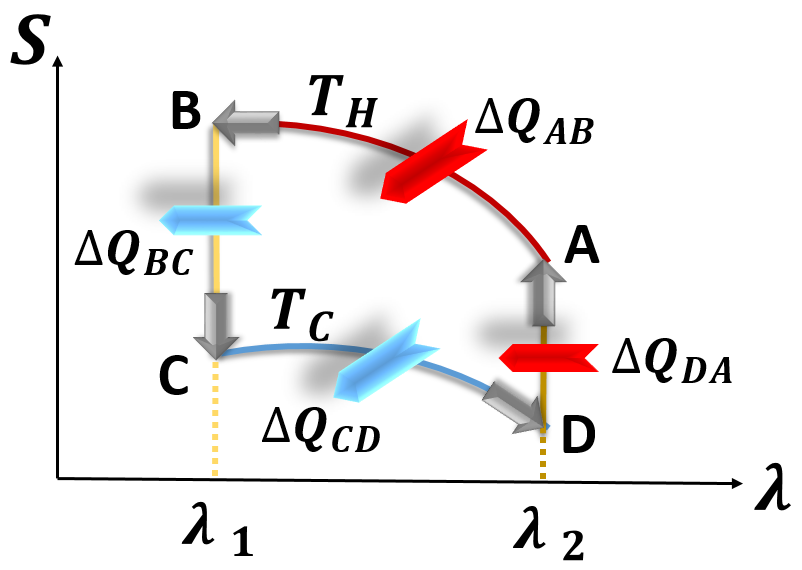}

\raggedright{}Figure. 1 (color online) Entropy-magnetic diagram ($S-\lambda$)
of the thermodynamic cycle based on the LMG model. Here, $\lambda_{1}$
and $\lambda_{2}$ are the magnetic fields in the two isomagnetic
field processes from B to C and from D to A. $T_{H}$ and $T_{C}$
are the temperatures of the two heat baths.
\end{figure}
\par\end{center}

In this section, we build a quantum heat engine with a spin system,
which is modeled as the LMG model. In the thermodynamic cycle (Fig.
1), the processes from A to B and from C to D are the quantum isothermal
processes, while the processes from B to C and from D to A are the
isomagnetic field processes.

During the isothermal process from A to B (C to D), the system is
kept in contact with a heat bath at temperature $T_{H}$ ($T_{C}$).
The external magnetic field, which is regarded as the generalized
coordinate in this thermodynamic cycle, is slowly reduced (increased)
from $\lambda_{2}$ ($\lambda_{1}$) to $\lambda_{1}$ ($\lambda_{2}$).
$\Delta Q_{AB}$ and $\Delta Q_{CD}$ represent the amounts of heat
exchange between the system and the heat baths during the two isothermal
processes, respectively. The working substance is always kept in thermal
equilibrium with the heat bath by assuming that the energy levels
of the system change much slower than the relaxation rates.

At the beginning of the isomagnetic field process from B to C (D to
A), the system is rapidly brought into thermal contact with the heat
bath at low temperature $T_{C}$ (high temperature $T_{H}$). No work
has been done during this process, because the external magnetic field
is fixed at $\lambda_{1}$ ($\lambda_{2}$) and the eigenenergies
of the working substance remain unchanged \cite{key-3}. The amounts
of heat exchange between the system and the heat baths $\Delta Q_{BC}$
($\Delta Q_{DA}$ ) is equal to the change in the internal energy
of the system.

Based on the definition of the entropy and the first law of thermodynamics,
we can calculate the amounts of the heat exchange in each thermodynamic
process as\begin{subequations}
\begin{eqnarray}
\Delta Q_{AB} & = & T_{H}\left(S_{B}-S_{A}\right),\label{appa}\\
\Delta Q_{BC} & = & U_{C}-U_{B},\label{appb}\\
\Delta Q_{CD} & = & T_{C}\left(S_{D}-S_{C}\right),\label{appc}\\
\Delta Q_{DA} & = & U_{A}-U_{D}.
\end{eqnarray}
\end{subequations} Here, $S_{\alpha}$ are the entropy of the system
in A, B, C, and D points in the cycle. The system's internal energy
$U_{\alpha}=\sum_{i}p_{i}\left(\lambda_{\alpha},T_{\alpha}\right)E_{i}\left(\lambda_{\alpha}\right)$
\cite{key-1}. As a result, the efficiency of the thermodynamic cycle
is

\begin{equation}
\eta=\frac{W}{Q_{H}}=\frac{\Delta Q_{AB}+\Delta Q_{BC}+\Delta Q_{CD}+\Delta Q_{DA}}{\Delta Q_{AB}+\Delta Q_{DA}},\label{eq:definatineta}
\end{equation}
where the work output per cycle $W=\Delta Q_{AB}+\Delta Q_{BC}+\Delta Q_{CD}+\Delta Q_{DA}$.
From Eq. (\ref{eq:definatineta}) and Fig. 1, one can find that the
net amount of heat input $Q_{H}$ into the the thermodynamic cycle
is determined by the heat transfer $\Delta Q_{AB}$ between the system
and the heat bath at temperature $T_{H}$ during the isothermal process
from A to B and that of the heat transfer $\Delta Q_{DA}$ in the
isomagnetic field process from D to A. For $N=2$ (or 20), $T_{H}=0.8$
(or 80), and $\eta_{C}=\left(T_{H}-T_{C}\right)/T_{H}=0.5$, we can
numerically calculate the efficiency of this heat engine from Eqs.
(\ref{eq:eigenenergy}), (\ref{eq:entropy}), and (\ref{eq:definatineta})
with $\lambda_{2}=4$ and $\lambda_{1}\in[0,4]$. The results are
shown in Figs. 2 (a) and (b).
\begin{figure}
\includegraphics[scale=0.27]{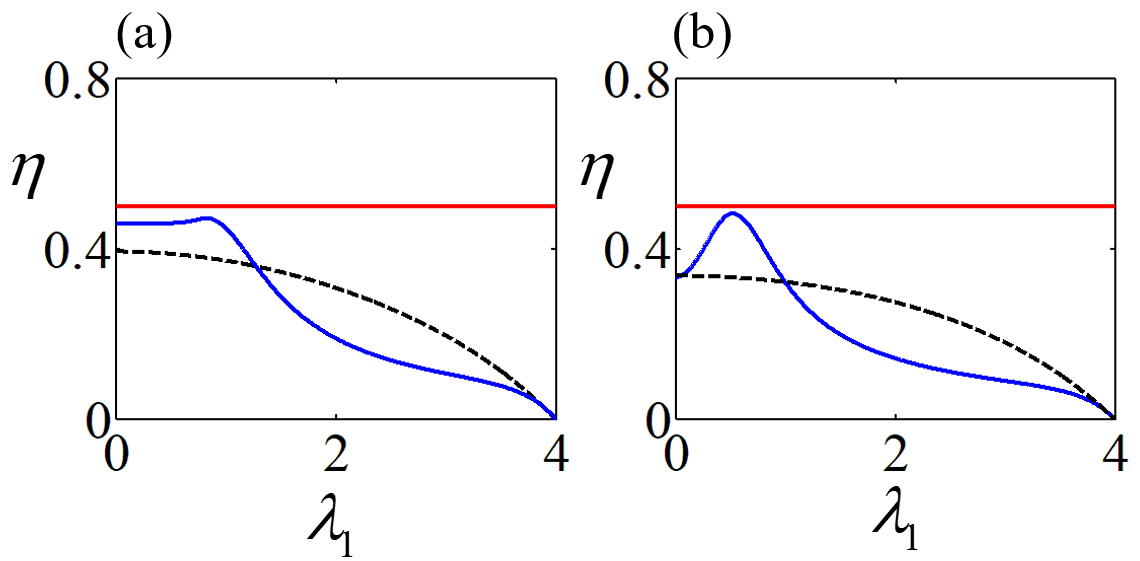}
\raggedright{}Figure. 2 (color online). Diagrams of the efficiency
$\eta$ as a function of $\lambda_{1}$ for the system with $N=20$
in (a) and $N=2$ in (b). The blue solid line and the black dotted
line correspond to the high temperature heat bath at $T_{H}=0.8$
and $T_{H}=80$, respectively. Here we guarantee that the Carnot efficiency
of each curve equals $0.5$, as indicated by the red solid line.
\end{figure}
It is found in Fig. 2(a) that if $T_{H}$ and $T_{C}$ are both very
low, the efficiency of the cycle is capable of approaching the Carnot
efficiency when $\lambda_{1}$ is below the QPT point of the LMG model.
In Fig. 2(b), we notice that when $N=2$, there is a maximum value
of efficiency at $\lambda_{1}\approx0.5$. According to Sec.II.B,
the crosspoint of the ground states' energy levels is just located
at $\lambda_{1}=1/2$. In the next Section, we will theoretically
and numerically prove that the positions of the maximum efficiency
of the quantum heat engine is always close to the energy level crosspoints.

\section{Maximum efficiency of the heat engine for system with finite $N$
\label{sec:Maximum-efficiency-of}}

\subsection{System with $N=2$ }

The eigenenergies for the LMG model with two interacting spins are
$E_{-1}=2\lambda-1$, $E_{0}=-2$, and $E_{1}=-2\lambda-1$. The crosspoint
of the ground state's energy levels is at $\lambda_{1}=1/2$. When
$\beta\gg1$ and $\lambda_{2}\gg1$, the population of the system
in the ground state is

\[
p\left(\lambda_{2},\beta\right)=\frac{e^{2\beta\lambda_{2}}}{e^{2\beta\lambda_{2}}+e^{-2\beta\lambda_{2}}+e^{\beta}}\approx1,
\]
which means that the system is mostly in its ground state. As a result,
$S_{A}=S\left(\lambda_{2},\beta_{H}\right)\approx0$, and $S_{D}=S\left(\lambda_{2},\beta_{C}\right)\approx0$
$\left(\beta_{C}>\beta_{H}\gg1\right)$. Therefore, the heat transfer
$\Delta Q_{DA}$ in the isomagenic process from D to A is written
as,

\begin{equation}
\Delta Q_{DA}=U_{A}-U_{D}=E_{G}^{\lambda_{2}}-E_{G}^{\lambda_{2}}=0.\label{eq:deltaQDA}
\end{equation}
In this case, the efficiency of the thermodynamic cycle is

\begin{equation}
\eta=\frac{T_{H}S_{B}+U_{C}-U_{B}-T_{C}S_{C}}{T_{H}S_{B}}.
\end{equation}
Notice that $S\left(\lambda_{\alpha},\beta_{\alpha}\right)=U_{\alpha}/T_{\alpha}+\ln Z_{\alpha}$,
the expression of the work can be simplified as $W=T_{H}\ln Z_{B}-T_{C}\ln Z_{C}$,
where the partition function is
\begin{equation}
Z\left(\lambda,\beta\right)=e^{\beta}\left[e^{-2\beta\lambda}+e^{2\beta\lambda}+e^{\beta}\right].
\end{equation}
When $\beta_{C}>\beta_{H}\gg1$ and $e^{-2\beta\lambda_{1}}\ll1(\lambda_{1}\neq0)$,
the partition functions at points B and C of the thermodynamic cycle
are

\begin{equation}
Z_{B}\approx e^{2\beta_{H}\left(2\lambda_{1}+1\right)}\left[1+e^{\beta_{H}\left(1-2\lambda_{1}\right)}\right],\label{eq:ZB}
\end{equation}
and

\begin{equation}
Z_{C}\approx e^{2\beta_{C}\left(2\lambda_{1}+1\right)}\left[1+e^{\beta_{C}\left(1-2\lambda_{1}\right)}\right],\label{eq:ZC}
\end{equation}
respectively. Combining Eqs. (\ref{eq:ZB}) and (\ref{eq:ZC}), we
see that

\[
W=\frac{\ln\left[1+e^{\beta_{H}\left(1-2\lambda_{1}\right)}\right]}{\beta_{H}}-\frac{\ln\left[1+e^{\beta_{C}\left(1-2\lambda_{1}\right)}\right]}{\beta_{C}}.
\]
Differentiating $W$ with respect to $\lambda_{1}$, one can write

\begin{equation}
\frac{\partial W}{\partial\lambda_{1}}=2\frac{e^{\beta_{C}\left(1-2\lambda_{1}\right)}-e^{\beta_{H}\left(1-2\lambda_{1}\right)}}{\left[1+e^{\beta_{H}\left(1-2\lambda_{1}\right)}\right]\left[1+e^{\beta_{C}\left(1-2\lambda_{1}\right)}\right]}.\label{eq:dW}
\end{equation}
As $\beta_{C}>\beta_{H}$, it is found that

\[
\begin{cases}
\frac{\partial W}{\partial\lambda_{1}}>0 & (\lambda_{1}<\frac{1}{2})\\
\frac{\partial W}{\partial\lambda_{1}}=0 & (\lambda_{1}=\frac{1}{2})\\
\frac{\partial W}{\partial\lambda_{1}}<0 & (\lambda_{1}>\frac{1}{2})
\end{cases}\text{.}
\]
This means that the work $W$ of this cycle has a maximum value at
$\lambda_{1}=1/2$, which is exactly the corsspoint of the energy
level of the ground states. Now we differentiate $Q_{H}$ with respect
to $\lambda_{1}$

\begin{equation}
\frac{\partial Q_{H}}{\partial\lambda_{1}}=T_{H}\frac{\partial S_{B}}{\partial\lambda_{1}}=\frac{\partial U_{B}}{\partial\lambda_{1}}+T_{H}\frac{\partial\left(\ln Z_{B}\right)}{\partial\lambda_{1}},\label{eq:differentiate QH}
\end{equation}
It follows from Eq. (\ref{eq:ZB}) that

\begin{equation}
U_{B}=-\frac{\partial\left(\ln Z_{B}\right)}{\partial\beta_{H}}=\frac{-2\lambda_{1}e^{2\beta_{H}\lambda_{1}}-e^{\beta_{H}}}{e^{2\beta_{H}\lambda_{1}}\left[1+e^{\beta_{H}\left(1-2\lambda_{1}\right)}\right]}-1,
\end{equation}
and

\begin{equation}
\frac{\partial Z_{B}}{\partial\lambda_{1}}=2\beta_{H}e^{\beta_{H}\left(2\lambda_{1}+1\right)}.
\end{equation}
Equation (\ref{eq:differentiate QH}) can now be further simplified
as

\begin{equation}
\frac{\partial Q_{H}}{\partial\lambda_{1}}=\frac{2\beta_{H}e^{2\beta_{H}\lambda_{1}}}{Z_{B}^{2}}e^{\beta_{H}}\left(1-2\lambda_{1}\right).\label{eq:dQ}
\end{equation}
With the help of Eq. (\ref{eq:definatineta}), one can differentiate
$\eta$ with respect to $\lambda_{1}$ as

\begin{equation}
\frac{\partial\eta}{\partial\lambda_{1}}=\frac{1}{Q_{H}^{2}}\left(\frac{\partial W}{\partial\lambda_{1}}Q_{H}-\frac{\partial Q_{H}}{\partial\lambda_{1}}W\right).\label{eq:deta}
\end{equation}
Combining Eqs. (\ref{eq:dW}), (\ref{eq:dQ}), (\ref{eq:deta}), and
taking $\lambda_{1}=1/2+\delta$ ($\delta\ll1$), we find that

\[
\left[\frac{\partial\eta}{\partial\lambda_{1}}\right]_{\lambda_{1}=\frac{1}{2}+\delta}=\frac{\beta_{H}^{2}}{\ln2}\left(2-\frac{T_{H}}{T_{C}}-\frac{T_{C}}{T_{H}}\right)\delta.
\]
Finally, we can conclude that

\begin{equation}
\mathrm{sgn}\left\{ \left[\frac{\partial\eta}{\partial\lambda_{1}}\right]_{\lambda_{1}=\frac{1}{2}+\delta}\right\} =-\mathrm{sgn}\left(\delta\right),\label{eq:efficiencysymbol}
\end{equation}
where $\mathrm{sgn}\left(x\right)\equiv x/\left|x\right|$ is the
sign function. Equation (\ref{eq:efficiencysymbol}) shows that the
efficiency of such a thermodynamic cycle has maximum value at $\lambda_{1}=1/2$
and $\eta\left(\lambda_{1}=1/2\right)=\eta_{Carnot}$. This result
can be checked by the numerical calculation directly, where we take
$N=2$, $T_{H}=0.6$, $T_{C}=0.3$, $\lambda_{1}\in[0,4]$, and $\lambda_{2}=4$.
The diagram of $\eta$ as a function of $\lambda_{1}$ is plotted
in Fig. 3(a), and the entropy of the system at the given temperature
as a function of $\lambda_{1}$ is showed in Fig. 3(b).
\begin{figure}
\includegraphics[scale=0.27]{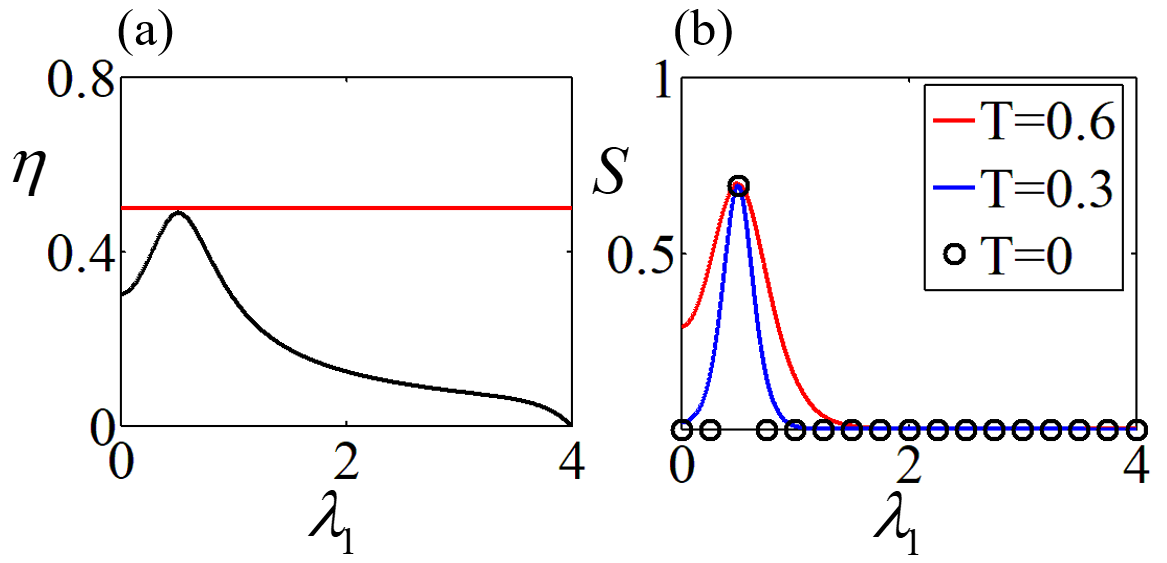}
\raggedright{}Figure. 3 (color online) (a) Diagram of the efficiency
$\eta$ as a function of $\lambda_{1}$ for the system with $N=2$.
$T_{H}=0.6$ is the temperature of the hot heat bath and $T_{C}=0.3$
is the temperature of the cold heat bath. The black solid line represents
the efficiency of the heat engine while the red solid line represents
the corresponding Carnot efficiency. (b) Diagram of the entropy $S$
as a function of $\lambda_{1}$, where the red solid line, the blue
solid line, and the black circle line represent the change in the
entropy when the temperatures ($T$) of the system are at $0.6$,
$0.3$, and $0$ respectively.
\end{figure}
It can be found in Fig. 3(a) that the efficiency takes the maximum
value at $\lambda_{1}=0.5$ and is equal to the Carnot efficiency.
In Fig. 3(b), one can see that when $\lambda_{1}=0.5$ , the blue
line is tangent to the red line, and the corresponding entropy value
is the same as the entropy of system at zero temperature represented
by the black circle. When $\lambda_{1}=1/2$, the ground states are
crossing with $U_{B}\approx U_{C}=-1$, and $S_{B}\approx S_{C}\approx\ln2$.
As a result,

\begin{equation}
\eta=\frac{T_{H}S_{B}-T_{C}S_{C}}{T_{H}S_{B}}\approx\eta_{C},
\end{equation}
which is also observed in Fig. 3(b). From another point of view, when
$\lambda_{1}$ is at the crosspoint of the ground states' energy levels,
the internal energy of point B is equal to that of point C. It implies
that there is no heat exchange between the system and the bath in
the process from B to C. On the other hand, Eq. (\ref{eq:deltaQDA})
indicates that the heat transfer in the process from D to A doesn't
exist ($\lambda_{2}\gg1$). There is no heat transfer in the two isomagnetic
processes of the thermodynamic cycle. Thus, from the entropy diagram,
the thermodynamic cycle we construct is similar to the Carnot cycle.
This is the fundamental reason why the efficiency of the quantum heat
engine we proposed can achieve the Carnot efficiency under this particular
case.

\subsection{System with $N>2$}

For $N>2$, the crosspoints are located at $\lambda_{1}\left(k\right)=(2k+1)/N$
$\left(k=0,1,\cdots N/2-1\right)$. Considering Eq. (\ref{eq:eigenenergy}),
we have

\[
E\left(M=k\right)=E\left(M=k+1\right)<E\left(M\neq k,k+1\right)<0.
\]
When $\beta\gg1$,

\[
e^{-\beta E\left(M=k\right)}=e^{-\beta E\left(M=k+1\right)}\gg e^{-\beta E\left(M\neq k,k+1\right)},
\]
the partition function

\[
Z=\sum_{M}e^{-\beta E(M)}\approx2e^{-\beta E\left(k\right)}.
\]
The entropy

\begin{align*}
S & =\beta U\left(\lambda_{1}\left(k\right),\beta\gg1\right)+\ln Z\left(\lambda_{1}\left(k\right),\beta\gg1\right)\\
 & \approx\beta E\left(k\right)+\ln\left[2e^{-\beta E\left(k\right)}\right]=\ln2,
\end{align*}
which is independent of the temperature. As a result,

\begin{equation}
\eta=\frac{T_{H}S\left(\lambda_{1},T_{H}\right)-T_{C}S\left(\lambda_{1},T_{C}\right)}{T_{H}S\left(\lambda_{1},T_{H}\right)}\approx\eta_{C}.\label{eq:peakvalue}
\end{equation}
It implies that when the temperature of the two heat bath are very
low and $\lambda_{1}$ takes the values for reaching the crosspoint
of the ground states's energy levels, the efficiency of the heat engine
approaches the Carnot efficiency. The numerical results of the efficiency
of the thermodynamic cycle with $N=4,6,8,$ and $10$ are given in
Fig. 4.
\begin{figure}
\includegraphics[scale=0.44]{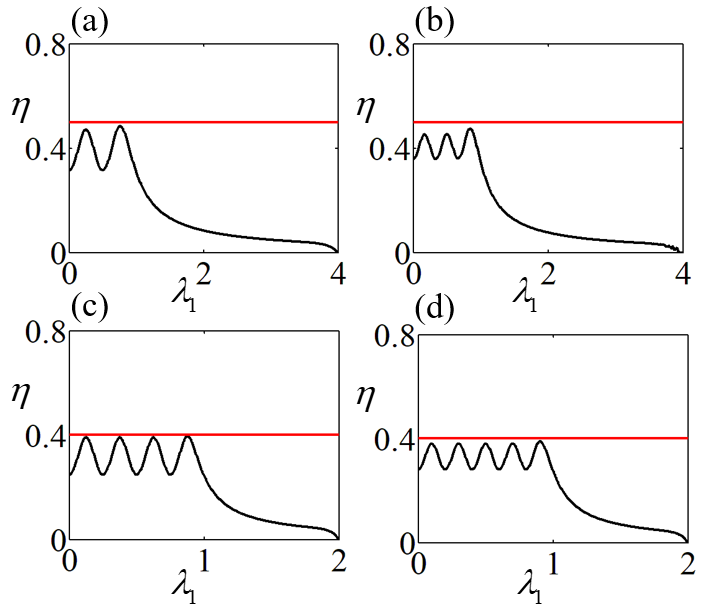}
\raggedright{}Figure. 4 (color online) Diagrams of the efficiency
$\eta$ as a function of $\lambda_{1}$, where the black solid lines
represents the efficiency of the heat engine while the red solid lines
represents the corresponding Carnot efficiency. (a) For the system
with $N=4$, the parameters $T_{H}=0.3$, $T_{C}=0.15$, $\lambda_{1}\in[0,4]$,
and $\lambda_{2}=4$. (b) For the system with $N=6$, the parameters
$T_{H}=0.2$, $T_{C}=0.1$, $\lambda_{1}\in[0,4]$, and $\lambda_{2}=4$.
(c) For the system with $N=8$, the parameters $T_{H}=0.1$, $T_{C}=0.06$,
$\lambda_{1}\in[0,2]$, and $\lambda_{2}=2$. (d) For the system with
$N=10$, the parameters $T_{H}=0.1$, $T_{C}=0.06$, $\lambda_{1}\in[0,2]$,
and $\lambda_{2}=2$.
\end{figure}
One can easily check that under different conditions, there are $N/2$
peaks in the diagram of $\eta$ varying $\lambda_{1}$, which is exactly
the same as the number of the crosspoints of the ground states' energy
level. The efficiency values corresponding to these peaks are also
consistent with the prediction in Eq. (\ref{eq:peakvalue}), which
is close to the Carnot efficiency $\eta_{C}$. These figures validate
our heretical analysis that the maximum value of the cycle's efficiency
is always obtained when $\lambda_{1}$ is at the crosspoints of energy
levels. It's obviously observed in these figures that there are $N/2$
peaks in the curve of efficiency for $N$-spin system. As $N$ increases,
the peaks gradually become flat and approach the red straight line
marked by the Carnot efficiency. As shown in Sec. IIB, the number
of the crosspoints of the ground states increases as $N$ increases.
Thus, we can make a reasonable speculation that the efficiency of
the thermodynamic cycle at the low temperature limit will always approach
the Carnot efficiency when $\lambda_{1}<\lambda_{c}$ and $N$ is
taken to be the thermodynamic limit ($N\rightarrow\infty$). This
will be proved analytically in the next section. In order to further
demonstrate the relationship between the cycle's efficiency and the
crosspoints of the energy levels, the plot of $\eta'=\partial\eta/\partial\lambda_{1}$
as a function of $\lambda_{1}$ under different situation are calculated
and illuminated in Fig. 5.
\begin{figure}
\includegraphics[scale=0.44]{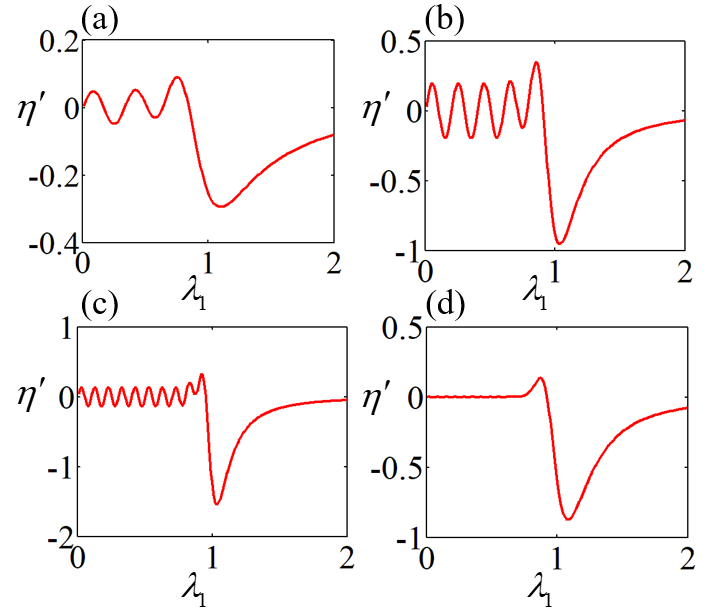}
\raggedright{}Figure. 5 (color online) Diagrams of $\eta'=\partial\eta/\left(\partial\lambda_{1}\right)$
as a function of $\lambda_{1}$, where $\lambda_{1}\in[0,2]$, and
$\lambda_{2}=2$. (a) For the system with $N=6$, the parameters $T_{H}=0.3$,
$T_{C}=0.2$. (b) For the system with $N=10$, the parameters $T_{H}=0.2$,
$T_{C}=0.1$. (c) For the system with $N=20$, the parameters $T_{H}=0.12$,
$T_{C}=0.06$. (d) For the system with $N=30$, the parameters $T_{H}=0.2$,
$T_{C}=0.1$.
\end{figure}
The diagrams show that there exists a valley around $\lambda_{c}=1$.
When $N$ is small, there are some fluctuations in the curves, which
are exactly corresponding to the crosspoints of the ground states'
energy levels. With the increase of $N$, the crosspoints of the ground
states' energy level of the system become more dense, and the fluctuations
diminish until they disappear.

\section{Efficiency of the heat engine under thermodynamic limit $N\rightarrow\infty$\label{sec:Efficiency-of-the}}

When $N$ is taken to be the thermodynamic limit ($N\rightarrow\infty$),
the partition function can be approximated as (details see Appendix)

\begin{equation}
Z=\frac{e^{\beta N\left(1+\lambda^{2}\right)/2}}{\sqrt{2N\beta}}\left[\mathrm{erf}\left(\frac{a+1}{K}\right)+\mathrm{erf}\left(\frac{-a}{K}\right)\right]\label{eq:partition function}
\end{equation}
where $\mathrm{erf}\left(x\right)=\sqrt{4/\pi}\int_{0}^{x}\exp\left(-\eta^{2}\right)d\eta$
is the Gauss error function, $a=-\left(1+\lambda\right)/2$, and $K=1/\sqrt{2N\beta}$.
With the help of Eq. (\ref{eq:partition function}), we can discuss
the efficiency of the heat engine under different circumstances.

\subsection{Efficiency at high temperature limit}

For $T\gg1$, $(a+1)/K,-a/K\ll1$

\[
\mathrm{erf}\left(\frac{a+1}{K}\right)\approx-\sqrt{\frac{2N\beta}{\pi}}(1-\lambda),
\]
and

\[
\mathrm{erf}\left(\frac{-a}{K}\right)\approx\sqrt{\frac{2N\beta}{\pi}}(1+\lambda).
\]
We have

\begin{equation}
Z\approx\frac{1}{\sqrt{\pi}}\exp\left[\frac{N\beta}{2}\left(1+\beta\lambda^{2}\right)\right],\label{eq:Z1}
\end{equation}
where the relation $\mathrm{erf}\left(\sqrt{2N}/2\right)\approx\mathrm{erf}\left(\infty\right)=1$
have been used. Taking the logarithm of both side of Eq. (\ref{eq:Z1}),
we have

\begin{equation}
\ln Z\approx\frac{1}{2}\ln\pi+\frac{N\beta}{2}\left(1+\beta\lambda^{2}\right).\label{eq:lz1}
\end{equation}
As a result, the internal energy

\begin{equation}
U=-\frac{\partial\left(\ln Z\right)}{\partial\beta}=-\left(\frac{N}{2}+N\beta\lambda^{2}\right)\text{.}\label{eq:u1}
\end{equation}
It follows from Eqs. (\ref{eq:definatineta}),(\ref{eq:lz1}), and
(\ref{eq:u1}) that

\[
\eta=\frac{\left(\lambda_{2}^{2}-\lambda_{1}^{2}\right)\left(\beta_{C}-\beta_{H}\right)}{\left(\lambda_{1}^{2}-\lambda_{2}^{2}\right)\beta_{H}+2\beta_{C}\lambda_{2}^{2}-2\beta_{H}\lambda_{1}^{2}}.
\]
By introducing $\kappa\equiv\lambda_{1}/\lambda_{2}\in[0,1]$, the
efficiency can be further simplified as

\begin{equation}
\eta=\frac{\left(1-\kappa^{2}\right)\eta_{C}}{1-\kappa^{2}+\eta_{C}\left(1+\kappa^{2}\right)},\label{eq:approximationeta1}
\end{equation}
which is independent of $N$. Especially, when $\kappa\ll1$, $\eta_{0}=\eta_{C}/\left(1+\eta_{C}\right)$.
For purposes of verifying the above analysis, we give the numerical
result of the efficiency as a function of $\lambda_{1}$ with $N=100$,
$\lambda_{1}\in\left[0,30\right]$, and $\lambda_{2}=30$, as shown
in Fig. 6
\begin{figure}
\includegraphics[scale=0.28]{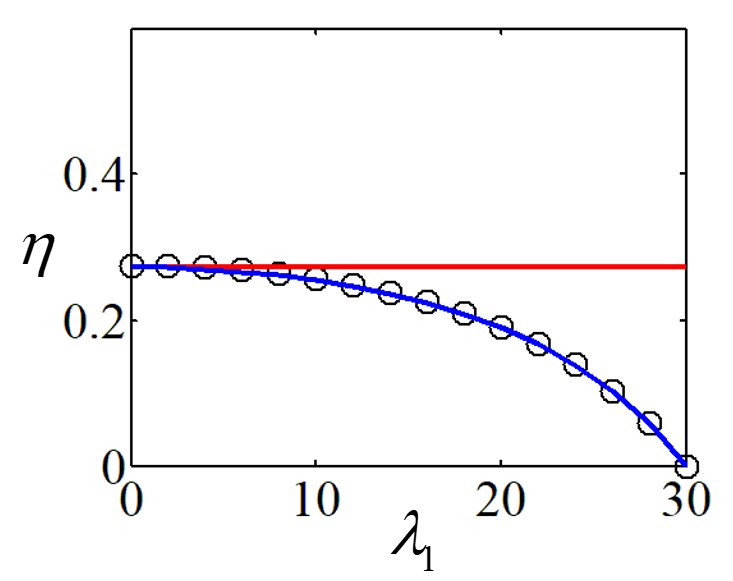}

\raggedright{}Figure. 6 (color online) Diagrams of $\eta$ as a function
of $\lambda_{1}$. Here, the black circle line is the numerical results,
the blue solid line is the analytical results of Eq. (\ref{eq:approximationeta1}),
and the red solid line is $\eta_{0}=\eta_{C}/\left(1+\eta_{C}\right)$,
where $\eta_{C}$ is the corresponding Carnot efficiency. The temperature
of the two heat baths are $T_{H}=800$ and $T_{C}=500$, respectively.
\end{figure}
. It can be seen in Fig. 6 that Eq. (\ref{eq:approximationeta1})
is consistent well with the numerical result, and when $\lambda_{1}\ll\lambda_{2}$,
the efficiency approaches $\eta_{0}=\eta_{C}/\left(1+\eta_{C}\right)$.

\subsection{Efficiency at low temperature limit}

In this section, we consider the working substance contacting with
the baths having temperature $T\ll1$. In the following, we will discuss
the partition function in two cases involving $\lambda<1$ and $\lambda>1$.
If $\lambda<1$

\[
\mathrm{erf}\left(\frac{a+1}{K}\right)+\mathrm{erf}\left(\frac{-a}{K}\right)\approx2\mathrm{erf}\left(\frac{\sqrt{2N\beta}}{2}\right)=2,
\]
such that the partition function is approximated as

\begin{equation}
Z=\frac{2}{\sqrt{2N\beta}}\exp\left[\frac{\beta N}{2}\left(1+\lambda^{2}\right)\right]\label{eq:Znew}
\end{equation}
The logarithm of Eq. (\ref{eq:Znew}) can be written as

\begin{equation}
\ln Z=\ln2-\frac{1}{2}\ln(2N\beta)+\frac{N\beta}{2}\left(1+\lambda^{2}\right)\approx\frac{N\beta}{2}\left(1+\lambda^{2}\right).\label{eq:lz2}
\end{equation}
As a result,

\begin{equation}
U=-\frac{\partial\left(\ln Z\right)}{\partial\beta}=-\frac{N}{2}\left(1+\lambda^{2}\right).\label{eq:u2}
\end{equation}
For $\lambda_{1}<\lambda_{2}<1$, it follows from Eqs. (\ref{eq:definatineta}),
(\ref{eq:lz2}), and (\ref{eq:u2}) that $\eta\approx0$. This is
checked by the numerical calculation, and the results are illstrated
in Figs. 7 (a) and (b).
\begin{figure}
\includegraphics[scale=0.28]{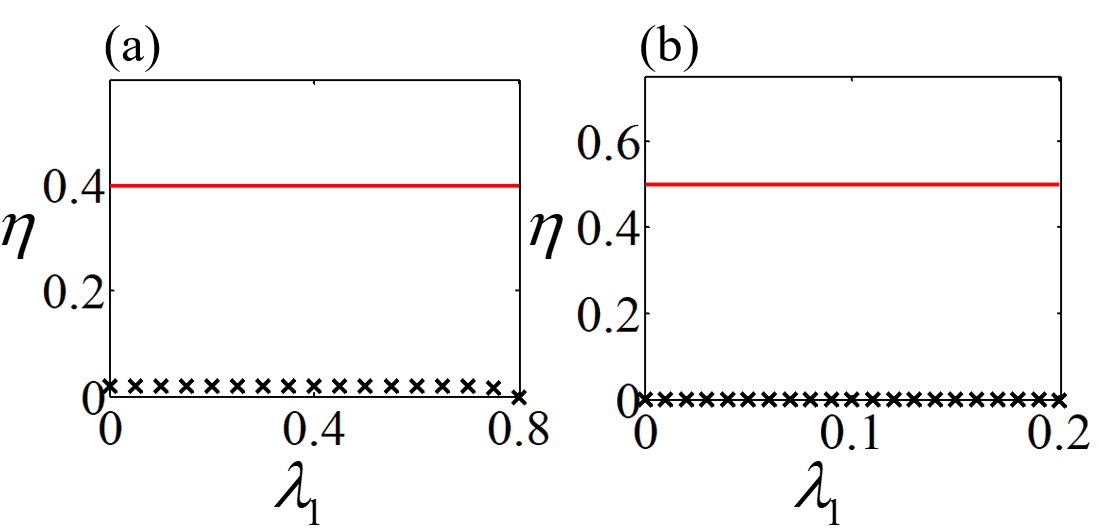}
\raggedright{}Figure. 7 (color online) Diagrams of the efficiency
$\eta$ as a function of $\lambda_{1}$, where the black fork lines
are the numerical results and the red solid lines represent the corresponding
Carnot efficiency. (a) For the system with $N=30$, the parameters
$T_{H}=0.5$, $T_{C}=0.3$, $\lambda_{1}\in[0,0.8]$, and $\lambda_{2}=0.8$.
(b) For the system with $N=50$, the parameters $T_{H}=0.2$, $T_{C}=0.1$,
$\lambda_{1}\in[0,0.2]$ and $\lambda_{2}=0.2$.
\end{figure}
On the other hand, when $\lambda\gg1$, one can write the energy spacing
of the LMG model's eigenenergies as
\[
\Delta E\left(M\right)=\left(\frac{4M}{N}-2\lambda\right)\Delta M=\left(\frac{4M}{N}-2\lambda\right),
\]
which is approximately equal to $2\lambda$. In this case, the system
would only stay in its ground state $\left|N/2,N/2\right\rangle $.
The reason is that the system can not be excited by the thermal energy
for $T\ll1$. The partition function is now given by $Z=\exp\left[-\beta E\left(N/2\right)\right]=\exp\left(\beta N\lambda\right).$
Taking the logarithm of both side, we have

\begin{equation}
\ln Z=\beta N\lambda,\label{eq:lz3}
\end{equation}
and the internal energy of the system is
\begin{figure}
\begin{centering}
\includegraphics[scale=0.27]{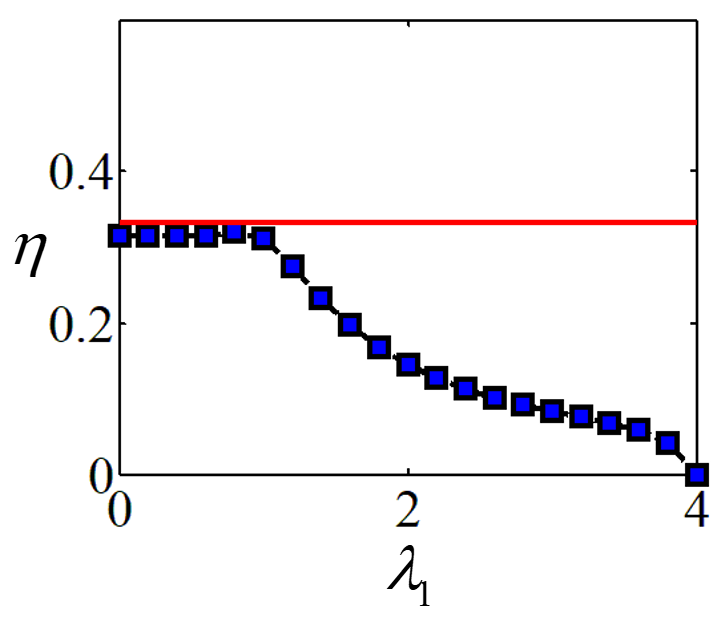}
\par\end{centering}
\raggedright{}Figure. 8 (color online) Diagram of $\eta$ as a function
of $\lambda_{1}$ for the system with $N=20$, where $\lambda_{1}\in[0,4]$,
and $\lambda_{2}=4$. $T_{H}=0.3$ is the temperature of the hot heat
bath. $T_{C}=0.2$ is the temperature of the cold heat bath. Here,
the blue square points are the numerical results and the red solid
line is the corresponding Carnot efficiency.
\end{figure}

\begin{equation}
U=-\frac{\partial\left(\ln Z\right)}{\partial\beta}=-N\lambda.\label{eq:u3}
\end{equation}
Combining Eqs. (\ref{eq:definatineta}), (\ref{eq:lz2}), (\ref{eq:u2}),
(\ref{eq:lz3}), (\ref{eq:u3}), and the condition $\lambda_{1}<1\ll\lambda_{2}$,
the efficiency of the thermodynamic cycle at low temperature limit
is obtained as $\eta=\eta_{C}$. This indicates that the efficiency
of the thermodynamic cycle is exactly the Carnot efficiency for system
with $N\gg1$. On the other hand, for $1\ll\lambda_{1}<\lambda_{2}$
and with the help of Eqs. (\ref{eq:lz3}) and (\ref{eq:u3}), one
has $\ln Z_{B}=N\beta_{H}\lambda_{1},\ln Z_{A}=N\beta_{H}\lambda_{2},\ln Z_{D}=N\beta_{C}\lambda_{2}$,
and $\ln Z_{C}=N\beta_{C}\lambda_{1}$. The work and the efficiency
are both equal to zero. We illustrate the numerical result in Fig.
8 to verify the above discussion, where the parameters we take are
$N=20$, $\lambda_{1}\in[0,4]$, and $\lambda_{2}=4$. It can be found
in Fig. 8 that the numerical results are consistent with the theoretical
analysis. When $\lambda_{1}<1$, the efficiency of the heat engine
is close to the Carnot efficiency. When $\lambda_{1}>1$, the efficiency
is rapidly reduced to zero.

\section{conclusion \label{sec:conclusion}}

In summary, we have studied quantum thermodynamic cycles with working
substance modeled as the Lipkin-Meshkov-Glick (LMG) model. It is found
that, for a finite system off thermodynamic limit, the efficiency
of the thermodynamic cycle can reach the Carnot limit at low temperature
when the the external magnetic field $\lambda_{1}$ corresponding
to one of the isomagnetic processes reaches the crosspoints of the
ground states' energy level. In the case of thermodynamic limit $\left(N\rightarrow\infty\right)$,
the analytical partition function of the LMG model in the thermal
equilibrium state is obtained and then used to calculate the efficiencies
of the heat engine at the high and low temperature limits. We show
that the efficiency of the quantum heat engine will achieve the Carnot
efficiency at low temperature, when the phases of the system in the
two isomagenetic processes are different. This observation implies
that the quantum phase transition (QPT) can improve the efficiency
of the thermodynamic cycle. In view of the fact that the LMG model
has been implemented on different experimental platforms, we expect
future quantum heat engines with the QPT to have the Carnot efficiency,
thereby achieving the purpose of using quantum techniques to enhance
the device efficiency.
\begin{acknowledgments}
This study is supported by the National Basic Research Program of
China (Grant No. 2014CB921403 \& No. 2016YFA0301201), the NSFC (Grant
No. 11421063 \& No. 11534002), and the NSAF (Grant No. U1530401).
\end{acknowledgments}

\appendix*

\section{Partition function}

The partition function is given by

\begin{equation}
Z=\sum_{M=-N/2}^{N/2}\exp\left[-\beta\left(\frac{2M^{2}}{N}-2\lambda M-\frac{N}{2}-1\right)\right].\label{eq:partition function-2}
\end{equation}
For system under thermodynamic limit $N\rightarrow\infty$, by taking
$x\equiv M/N$, the summation in Eq. (\ref{eq:partition function-2})
is approximated as an integral $\sum_{M=-N/2}^{M=N/2}=N\int_{-1/2}^{1/2}dx$.
The partition function can be rewritten as $Z=\exp\left(\beta N/2\right)\int_{-1/2}^{1/2}\mathrm{exp}\left[-2N\beta(x^{2}-\lambda x)\right]dx$.
After some algebraic manipulation, one has

\begin{align}
Z & \approx\exp\left[\frac{N\beta}{2}\left(1+\lambda^{2}\right)\right]\int_{-1/2}^{1/2}e^{-2N\beta\left(x-\frac{\lambda}{2}\right)^{2}}dx\\
 & =\exp\left[\frac{N\beta}{2}\left(1+\lambda^{2}\right)\right]K\int_{a/K}^{(a+1)/K}e^{-y^{2}}dy
\end{align}
Here, $a\equiv-\left(1+\lambda\right)/2$ and $K\equiv1/\sqrt{2N\beta}$.
Finally, with the help of the Gauss error function $\mathrm{erf}\left(x\right)=\sqrt{4/\pi}\int_{0}^{x}\exp\left(-\eta^{2}\right)d\eta$,
the partition function is obtained as

\begin{equation}
Z=\frac{e^{\beta N\left(1+\lambda^{2}\right)/2}}{\sqrt{2N\beta}}\left[\mathrm{erf}\left(\frac{a+1}{K}\right)+\mathrm{erf}\left(\frac{-a}{K}\right)\right]
\end{equation}

\end{document}